\documentclass[aps,prl,reprint, superscriptaddress]{revtex4-2}

\usepackage{siunitx}
\usepackage{cancel}
\usepackage{graphicx}
\usepackage{dcolumn}
\usepackage{bm}
\usepackage[utf8]{inputenc}
\usepackage[T1]{fontenc}
\usepackage{etoolbox}
\usepackage[caption=false]{subfig}
\usepackage{newtxtext,newtxmath}
\RequirePackage{fix-cm}
\usepackage[normalem]{ulem}
\usepackage{hyperref}
\hypersetup{
    colorlinks=true,
    linkcolor=blue,
    filecolor=magenta,      
    urlcolor=cyan,
}
\usepackage[svgnames]{xcolor}
\usepackage{bbm}
\usepackage{braket}
\usepackage{qcircuit}
\usepackage{soul}

\DeclareSIUnit\bar{bar}
\DeclareSIUnit{\belmilliwatt}{Bm}
\DeclareSIUnit{\dBm}{\deci\belmilliwatt}
\DeclareSIUnit\bar{bar}
\begin{document}

\title{Frequency Comb as a Probe of Fluctuation Mechanisms in Nanomechanical Systems} 

\author{Maria Kallergi}
\affiliation{Department of Electrical Engineering, School of Computation, Information and Technology,
Technical University of Munich, 85748 Garching, Germany}

\author{Daniel K.~J. Bone{\ss}}
\affiliation{Department of Physics, University of Konstanz, 78457 Konstanz, Germany}
\affiliation{Institute for Complex Quantum Systems and Center for Integrated Quantum Science and Technology,
Ulm University, Albert-Einstein-Allee 11, 89069 Ulm, Germany}

\author{Maximilian Seitner}
\affiliation{Department of Physics, University of Konstanz, 78457 Konstanz, Germany}

\author{Wolfgang Belzig}
\affiliation{Department of Physics, University of Konstanz, 78457 Konstanz, Germany}

\author{Mark I. Dykman}
\affiliation{Department of Physics and Astronomy, Michigan State University, East Lansing, Michigan 48824, USA}

\author{Eva M. Weig}
\affiliation{Department of Electrical Engineering, School of Computation, Information and Technology,
Technical University of Munich, 85748 Garching, Germany}
\affiliation{Munich Center for Quantum Science and Technology (MCQST), 80799 Munich, Germany}
\affiliation{TUM Center for Quantum Engineering (ZQE), 85748 Garching, Germany}

\date{\today}

\begin{abstract}
Frequency combs are indispensable in metrology. We show that a frequency comb also probes frequency fluctuations of the system. In our nanomechanical resonator, the comb arises from self-sustained vibrations of a resonantly driven nonlinear mode. The comb lineshapes reveal exponentially correlated (Ornstein--Uhlenbeck) frequency noise. Its correlation time agrees with the independently measured relaxation time of the vibrations. The noise leads to anomalous phase diffusion. Its mechanism is generic: like the comb itself, it stems from the strong nonlinearity of the self-sustained vibrations.
\end{abstract}


\maketitle 

 \textit{Introduction} --- Frequency fluctuations in vibrational systems are of major interest for physics and applications. Because of time-translation symmetry, the effect of frequency fluctuations on self-sustained vibrations, from lasers to atomic, electronic, and mechanical clocks, accumulates over time in the vibration phase. This limits the precision of clocks or, equivalently, broadens the vibrational spectrum~\cite{Berstein1938,Rytov1956,Lax1967,Demir2000,Zhang2015}, the nonzero laser linewidth being an example~\cite{Schawlow1958,Exter1991}. Understanding the mechanisms of frequency and the resulting phase fluctuations remains a major challenge that has been attracting much attention across various fields~\cite{Henry1982,Hajimiri1998,Fong2012,Loh2013,Moser2014,Ludlow2015,Sansa2016,Maillet2018,Bachtold2022,Kanellopulos2025,Iachkula2026}.
 
Frequency combs can provide insight into frequency fluctuations. They present equidistant peaks in the spectrum of a vibrational system and therefore play a fundamental role in metrology and sensing, a prominent example being optical atomic clocks~\cite{Diddams1999,Fortier2019,Udem2002,Bisht2026}. Combs often arise from the vibration nonlinearity. The spectra of the individual comb lines are then determined by the frequency fluctuations of the underlying generating system. Because there are many lines, their spectra provide a highly sensitive probe of the fluctuation mechanism.

In this paper, we show that the strong nonlinearity underlying the comb formation also shapes the frequency fluctuations. Accumulated in the phase, they give rise to pronounced anomalous phase diffusion.
The frequency fluctuations primarily come from fluctuations of the amplitude of the self-sustained nonlinear vibrations. A major source of amplitude fluctuations is broadband thermal noise~\cite{Albrecht1991,Cleland2002,Cleland2005,Maillet2018,Huang2019,Demir2021,Miller2021,Villanueva2023}. When transferred to the vibration frequency, the noise is filtered since the amplitude typically has a relaxation time far longer than the vibration period. The resulting filtered white noise with exponential correlations is known as Ornstein--Uhlenbeck (OU) noise. Consequently, phase diffusion displays a characteristic time dependence, and the comb lines are strongly non-Lorentzian.

We perform high-resolution measurements of a frequency comb on a nanomechanical system. Its lowest-frequency eigenmode is resonantly driven into forced vibrations at the drive frequency $\omega_\mathrm{d}$. When the drive becomes sufficiently strong, these vibrations become unstable, and the mode starts oscillating about the state of forced vibrations~\cite{Ochs2022} at a frequency much smaller than $\omega_\mathrm{d}$. In the frame rotating at $\omega_\mathrm{d}$, the self-sustained oscillations are strongly non-sinusoidal.

In the laboratory frame, these slow non-sinusoidal oscillations modulate the forced vibrations and produce equidistant peaks in the mode's spectrum, a phononic frequency comb.
We unambiguously trace the shapes and widths of all observed comb lines to OU frequency noise.
 
In a broader context, frequency combs  present a highly sensitive means to explore spectral features. Thanks to the multitude of the comb lines, they are ideally suited to reveal the mechanisms of the underlying fluctuations. We expect this generic approach to equally apply to other types and implementations of frequency combs.

\begin{figure}[t!]
    \centering
    \includegraphics[]{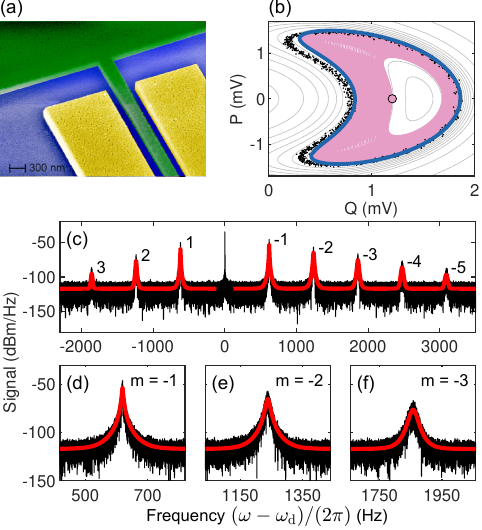}
    \caption{(a)~False-color scanning electron micrograph of a doubly-clamped silicon nitride string resonator (green) and adjacent electrodes (gold). (b)~Pink line depicts a theoretical trajectory toward the stable limit cycle in the rotating frame in the weak damping limit. The starting point is shown by an open circle. The stable limit cycle is shown by the blue trajectory. Gray lines show the contours of $g(Q,P)$. The measured limit cycle is shown as black dots for a drive power of \SI{0}{\dBm} and a drive frequency detuned from resonance by \SI{840}{\hertz}.  (c)~Phononic frequency comb generated from a single driven nonlinear nanomechanical mode for the same drive parameters, including measured data (black) and global fit of comb lines with Ornstein--Uhlenbeck (OU) noise (red). (d)--(f)~Close-ups of the lines $m=-1,-2,-3$ along with the OU fit (see Supplemental Material~\cite{Note1} for close-ups of further comb lines).
    }
   
    \label{fig:semduf}
\end{figure}

\textit{System and setup} ---
The nanomechanical resonator investigated in this work is a doubly-clamped string as depicted in the false-color scanning electron micrograph of Fig.~\ref{fig:semduf}(a). 
It is fabricated from strongly pre-stressed stoichiometric silicon nitride on a fused silica substrate, enabling large, dissipation-diluted quality factors, and features a width of \SI{270}{\nano\metre}, a thickness of \SI{100}{\nano\metre} and a length of \SI{55}{\micro\metre}.

The two flanking electrodes adjacent to the string enable dielectric drive through the Kelvin polarization force, while a heterodyne microwave-cavity-assisted detection scheme is used for the readout of the mechanical vibration~\cite{Unterreithmeier2009,Faust2012}. The cavity is driven on resonance at \SI{3.6}{\giga\hertz} to exclude backaction effects while maximizing the detection sensitivity. Application of a DC voltage allows frequency tuning of the fundamental out-of-plane mode of the nanostring. We keep it fixed at \SI{5}{\volt} to preclude coupling to other flexural modes, so that only a single mode is excited. All experiments are performed at room temperature (\SI{295}{\kelvin}) and in high vacuum at about \SI{e-4}{\milli\bar}, using an HF2LI lock-in amplifier (Zurich Instruments).

At weak drive powers, the nanomechanical resonator behaves as a harmonic oscillator, and its Lorentzian response yields the bare eigenfrequency $\omega_0/(2\pi) = \SI{6.519}{\mega\hertz}$ and the linewidth $2\Gamma_0/(2\pi) = \SI{21}{\hertz}$, where $\Gamma_0$ is defined as the amplitude decay rate. The resonator is thus strongly underdamped in the laboratory frame, with a quality factor of approx. $3 \times 10^5$.
As the drive power increases, the frequency response becomes asymmetric, indicating the onset of a Duffing nonlinearity. At even stronger drive, we enter the beyond-Duffing regime, where higher-order coefficients are needed to describe the response; Ref.~\cite{Kallergi2026} details how we accurately determine them.

In the laboratory frame, the mode coordinate $q(t)$ obeys the equation of motion
\begin{equation}
    \ddot{q} + 2\Gamma(q,\dot{q},F)\,\dot{q} + \omega_0^2 q + \beta q^2 + \gamma q^3 = \frac{F}{M} \cos(\omega_\mathrm{d} t) + \xi(t),
    \label{eom}
\end{equation}
where $\Gamma(q,\dot{q},F)$ comprises the linear and nonlinear damping as well as the resonantly induced friction force (RIFF) responsible for the comb generation~\cite{Ochs2022,Dykman2019}. The quadratic and cubic nonlinearities are denoted by $\beta$ and $\gamma$, respectively, $M$ is the effective mass, $F$ and $\omega_\mathrm{d}$ are the drive amplitude and frequency, and $\xi(t)$ is the thermal noise.

We describe the dynamics in the frame rotating at $\omega_\mathrm{d}$ by introducing the slow variables
\begin{equation}
    Q + iP = \left( q + \frac{ip}{M\omega_\mathrm{d}} \right) e^{i\omega_\mathrm{d} t},
\end{equation}
where $p = M\dot{q}$ is the momentum in the laboratory frame, and $Q$ and $P$ are the in-phase and quadrature components of the vibration in phase space. The rotating-wave approximation (RWA) averages over the fast oscillations at $\omega_\mathrm{d}$.

Within the RWA, the conservative dynamics are governed by the effective Hamiltonian
\begin{align}
    g(Q, P) = &-\frac{1}{2}\delta\omega\,(Q^2 + P^2) + \frac{M\omega_\mathrm{d}}{8}\alpha_1 (Q^2 + P^2)^2 \nonumber \\
    &- \frac{F}{2M\omega_\mathrm{d}}Q,
    \label{eq:heff_main}
\end{align}
for a detuning $\delta\omega = \omega_\mathrm{d} - \omega_0$. The nonlinear coefficient $\alpha_1$ is determined primarily by the Duffing nonlinearity $\gamma$~\cite{Kallergi2026}. Higher-order nonlinear coefficients enter as additional terms in powers of $Q^2 + P^2$. A detailed theoretical description is given in the Supplemental Material~\footnote{See Supplemental Material at [URL will be inserted by publisher] for detailed theoretical analysis, further information on the fitting procedures, and additional experimental data.}.

Figure~\ref{fig:semduf}(b) shows the dynamics in the quasienergy landscape $g(Q,P)$ for a detuning $\delta\omega/(2\pi) = \SI{840}{\hertz}$ and a drive power of \SI{0}{\dBm}. 
We plot the theoretical evolution (pink trace) from an initial condition (pink open circle) towards the stable limit cycle (blue trace). It includes the RIFF and the first three nonlinear coefficients of $g(Q,P)$; Ref.~\cite{Note1} details the model and lists all parameters. The resulting limit cycle is profoundly non-elliptical, which is a signature of non-sinusoidal vibration. We overlay the measured limit-cycle trajectory for the same parameters (black dots), calibrated as described in Ref.~\cite{Kallergi2026}. Theory and experiment are in excellent agreement.

\textit{Frequency comb with phase diffusion} --- 
Figure~\ref{fig:semduf}(c) shows the measured power spectral density for the same parameters as in Fig.~\ref{fig:semduf}(b), featuring the frequency comb. In the spectrum, the non-sinusoidal vibrations in the rotating frame appear as equidistant peaks around the drive frequency, spaced by the limit-cycle frequency $\nu_\mathrm{c}/(2\pi) = \SI{618.9}{\hertz}$. 

This frequency is well-defined because the vibrations are underdamped also in the rotating frame, $2\Gamma \ll \nu_\mathrm{c}$.
Without noise, the motion along the limit cycle is then periodic with phase $\psi(t) = \nu_\mathrm{c} t$, and the comb lines are $\delta$-peaks at $\omega_\mathrm{d} - m\nu_\mathrm{c}$~\cite{Ochs2022}.
The measured lines, however, have a finite width well above the \SI{30}{\milli\hertz} spectral resolution of the experiment (see Figs.~\ref{fig:semduf}(d)--(f) and~\ref{fig:remsat} for close-ups of the comb lines in Fig.~\ref{fig:semduf}(c)).

In the presence of frequency noise $\zeta(t)$, the phase obeys
\begin{equation}
    \dot{\psi} = \nu_\mathrm{c} + \zeta(t).
    \label{eq:phase}
\end{equation}
For stationary Gaussian fluctuations, the power spectrum becomes
\begin{widetext}
\begin{equation}
    S(\omega) = \frac{1}{2}\sum_m |z_m|^2 \int_0^\infty d\tau\, \cos\left[(\omega - \omega_\mathrm{d} + m\nu_\mathrm{c})\tau\right] \exp\left[-m^2 \int_0^\tau \Xi(x)(\tau - x)\,dx\right],
    \label{eq:psd_ksi}
\end{equation}
\end{widetext}
where $m$ is the comb line index, and $z_m = Q_m + i P_m$ are the Fourier components of the self-sustained vibrations. The function $\Xi(t) = \langle \zeta(t)\zeta(0) \rangle$ is the noise correlation function. It determines the lineshape of the comb lines and, through the factor $m^2$, how the linewidths grow with comb line number.

{\it Nature of the frequency noise}---The most common model of frequency noise is white Gaussian noise, cf.~\cite{Berstein1938,Cleland2002,Fong2012,Sansa2016,Miller2021}. It yields Lorentzian comb lines with widths growing as $m^2$. 
Another frequently encountered type is $1/f^\alpha$ noise, which can be modeled as Gaussian but not white. It has been observed with $\alpha \le 1$  in nano-~\cite{Cleland2002,Fong2012,Maillet2018,Roukes2026} and micromechanical systems~\cite{Sun2016}. The effect of $1/f$ noise on the spectrum of a nonlinear resonator was theoretically investigated in Refs.~\cite{Demir2002,Kenig2012}, and  frequency fluctuations due to low-frequency fluctuators were recently reported in nanomechanical phononic systems~\cite{Yuksel2026,Maksymowych2025a,Hitchcock2026}.

A different model of noise that has been broadly used in various contexts, from the theory of rare events to stochastic resonance to quantum optics, cf.~\cite{Bray1989,Gammaitoni1998,Carmele2020}, is exponentially correlated noise described in the seminal paper of Uhlenbeck and Ornstein~\cite{Uhlenbeck1930}. This OU noise is low-pass filtered white Gaussian noise, its power spectrum is Lorentzian. The three noise models are discussed in more detail in Ref.~\cite{Note1}.

For our phononic frequency comb, the mechanism outlined in the introduction predicts OU noise: Thermal noise drives fluctuations of the quasienergy $g$ of the limit cycle, which relax over a finite time. Since $\nu_\mathrm{c}$ depends on $g$, these fluctuations translate into frequency fluctuations with the noise correlation function
\begin{equation}
    \Xi(t) =\frac{\mathcal{D}}{\delta}e^{-|t|/\delta},
    \label{expcor}
\end{equation}
where  $\delta$ is the noise  correlation time and $\mathcal{D}\propto (\partial\nu_\mathrm{c}/\partial g)^2$ is the noise strength. Inserting Eq.~\eqref{expcor} into Eq.~\eqref{eq:psd_ksi}, the exponent becomes $-m^{2}\mathcal{D}[\tau+(e^{-\tau/\delta}-1)\delta]$. 

We fit the OU-determined lineshape to all comb lines simultaneously, with $\mathcal{D}$ and $\delta$ as global fit parameters. The fit, included as a red line in Figs.~\ref{fig:semduf}(c)--(f) and~\ref{fig:remsat}, yields 
$\mathcal{D} = \SI{8.0 \pm 0.3}{\radian\squared\per\second}$ and $\delta = \SI{29.5 \pm 0.9}{\milli\second}$. 
The fitted lineshapes (red) show excellent agreement with the measured comb lines (black). Fits with white or $1/f^\alpha$ noise, in contrast, fail to describe the lineshapes~\cite{Note1}. This unambiguously identifies OU frequency noise as the source of phase diffusion. To the best of our knowledge, this is the first direct observation of OU frequency noise in a nanomechanical resonator.

\begin{figure}[]
    \centering
    \includegraphics[]{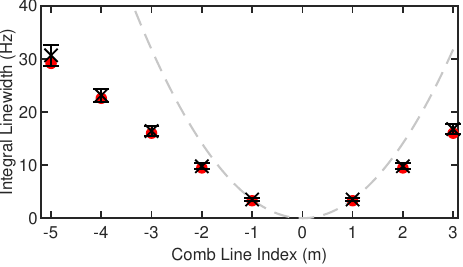}
    \caption{Comb linewidths. Integral linewidths of the comb lineshapes versus comb line number $m$, extracted from the measured spectra (black) and calculated from the fitted OU model (red). The two coincide within the experimental uncertainty. White Gaussian noise would instead lead to a quadratic growth (gray, $\propto m^2$, scaled to $m = \pm 1$).
    }
      \label{fig:linewidth}
\end{figure}

\textit{Comb linewidths} --- 
The correct description of the lineshapes allows us to extract the linewidths of all comb lines. We define the linewidth of each line as its integral width, i.e., the spectral area divided by the peak height. For the theory, we evaluate Eq.~\eqref{eq:psd_ksi} with the OU correlation function, Eq.~\eqref{expcor}, for the parameters obtained from the global fit. 
For the data, we use the same definition, but extracting the spectral area requires care because the tails of the comb lines merge with the noise floor. The End Matter describes how we approximate the tails to correctly estimate the noise floor.
Figure~\ref{fig:linewidth} shows the resulting linewidths as a function of comb line number $m$. Measured (black) and theoretical (red) linewidths coincide within the experimental uncertainty. They grow with $|m|$, but much more slowly than the parabola expected for white Gaussian noise (gray dashed line). Scaled to the linewidth at $m = \pm 1$, the parabola deviates strongly from the data for all higher comb lines.

\textit{Noise origin} ---
Frequency fluctuations of a self-sustained oscillator have two sources~\cite{Bachtold2022}: noise acting directly on the phase $\psi$, and noise acting on the quasienergy $g$, which the nonlinearity converts into fluctuations of $\nu_\mathrm{c}$. Direct phase noise comes primarily from white frequency noise and thus yields Lorentzian comb lines with widths growing as $m^2$,  which is incompatible with our data. The strong nonlinearity, evident from the profoundly non-elliptical limit cycle in Fig.~\ref{fig:semduf}(b) and the many comb lines in Fig.~\ref{fig:semduf}(c), makes the conversion of quasienergy fluctuations a dominant mechanism. The correlation time of these fluctuations sets the correlation time of the resulting frequency noise and makes it of the OU type~\cite{Miller2021}, as observed.

\begin{figure}[]
    \centering
    \includegraphics[]{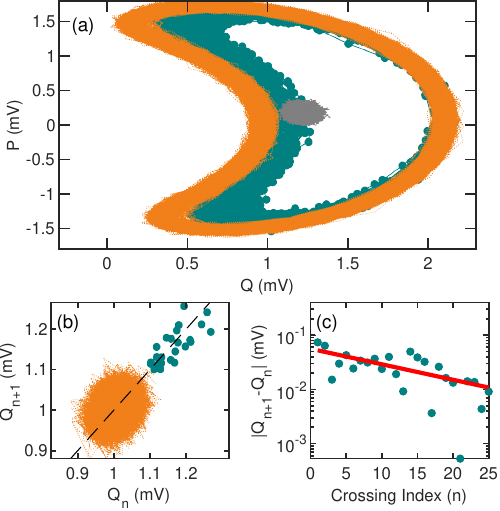}
    \caption{Formation of the limit cycle. (a) Phase space trajectory where the detuning of the drive (\SI{0}{\dBm}) is abruptly switched from $\delta\omega/(2\pi) = 0$ to \SI{840}{\hertz}. The system leaves the now unstable fixed point (gray), evolves towards the limit cycle (teal), and finally settles on it (orange).
    (b) Poincar\'e map $Q_{n+1}$ versus $Q_n$ of the crossings of the trajectory through $P = 0$ with $\dot{P} > 0$. The teal points show the approach to the limit cycle, and the orange points the noise-broadened stable limit cycle. The dashed line represents the $Q_{n+1} = Q_n$ line.
    (c) Step $|Q_{n+1} - Q_n|$ between consecutive crossings versus crossing number $n$ in the transient regime (teal). The exponential fit (red) yields the Floquet exponent $\lambda$ corresponding to a relaxation time $\delta = \SI{25 \pm 6}{\milli\second}$, in agreement with the OU correlation time from the comb lineshapes.}
    \label{fig:per}
\end{figure}

\textit{Limit-cycle formation} ---
Our analysis attributes the correlation time $\delta$ of the frequency noise to the relaxation of the quasienergy toward the limit cycle. 
We measure this relaxation directly by recording how the system approaches the limit cycle after the drive is tuned across the supercritical Hopf bifurcation~\cite{Ochs2022}. Since this measurement does not rely on noise, it provides an independent determination of $\delta$.
We start with the drive at \SI{0}{\dBm} at zero detuning, where the system rests at a stable fixed point in the rotating frame, and abruptly switch the drive frequency to $\delta\omega/(2\pi) = \SI{840}{\hertz}$, where the previously stable fixed point becomes unstable and is replaced by a stable limit cycle.
Both quadratures recorded over time are displayed as a phase space plot in Fig.~\ref{fig:per}(a). We observe the system leaving the fixed point (gray), evolving towards (teal) and settling on the limit cycle (orange). 
To quantify the relaxation time of the system towards the limit cycle, we record the crossings of the trajectory through $P = 0$ at the smaller of the two $Q$ values per revolution, i.e., the one with $\dot{P} > 0$, and plot the Poincar\'e map $Q_{n+1}$ versus $Q_n$ in Fig.~\ref{fig:per}(b).

The crossings move along the diagonal (teal) towards a dense cluster (orange), which marks the limit cycle. 
The step $|Q_{n+1} - Q_n|$ between consecutive crossings is depicted in Fig.~\ref{fig:per}(c). It decays exponentially with crossing number $n$, and a fit $\propto e^{-\lambda n}$ (red line) yields the Floquet exponent $\lambda = 0.065(16)$.

The frequency of the established limit cycle is $\nu_\mathrm{c}/(2\pi) = \SI{611.7}{\hertz}$. This results in a period of $\tau_\mathrm{c} = 2\pi/\nu_\mathrm{c} = \SI{1.63}{\milli\second}$, giving the relaxation time $\delta = \tau_\mathrm{c}/\lambda = \SI{25 \pm 6}{\milli\second}$.
This agrees within its uncertainty with the correlation time $\delta = \SI{25.4 \pm 0.5}{\milli\second}$ of the OU noise obtained from the global lineshape fit (see \cite{Note1} for more details of the measurements). This measurement was taken months after the data in Fig.~\ref{fig:semduf}, which explains the slight change in the limit-cycle frequency and noise correlation time.

\textit{Conclusion} ---
We use the spectra of the phononic  frequency comb lines as a probe of the mechanism of frequency fluctuations of the source of the comb. The high sensitivity of the method stems from the multiplicity of the comb lines. In our system, the comb emerges as a result of nonsinusoidal self-sustained vibrations in the rotating frame of a resonantly driven  single nanomechanical mode. We find that the lineshapes of all comb lines are determined by phase diffusion due to an  exponentially correlated frequency noise.
The noise has the same origin as the comb itself: both come from the strong nonlinearity of the self-sustained vibrations. In particular, this nonlinearity converts thermal-noise induced quasienergy fluctuations  into exponentially correlated frequency fluctuations. An independent and direct measurement of the  time of relaxation toward the state of self-sustained vibrations agrees with the noise correlation time, confirming our interpretation.  The fluctuation mechanism is therefore generic. The results provide a new insight into the origin of frequency fluctuations that ultimately limit the precision of comb-based measurements and suggest new strategies for improving comb technology.

\section{Acknowledgments}
M.K., D.K.J.B., W.B., and E.M.W. gratefully acknowledge financial support from the Deutsche Forschungsgemeinschaft (DFG, German Research Foundation) through Project-ID No.~425217212-SFB 1432 and under Germany’s Excellence Strategy—EXC-2111—390814868. The research is further supported by the Bavarian state government with funds from the Hightech Agenda Bavaria. M.I.D. gratefully acknowledges partial support from the Gordon and Betty Moore Foundation (Award No. GBMF12214) and the US Defense Advanced Research Projects Agency (Grant No. HR0011-23-2-004). D.K.J.B. acknowledges funding by the Carl-Zeiss-Foundation through the Carl-Zeiss-Foundation Center QPhoton.

\section{Data Availability}
Experimental data are available at 10.5281/zenodo.xxxxxxx.

\section{End Matter} 
\appendix
\textit{Spectral resolution}---Resolving the lineshapes of the comb lines requires a spectral resolution well below their linewidths. We record the spectra with an HF2LI lock-in amplifier (Zurich Instruments), which demodulates the signal at the drive frequency, thereby shifting the comb to baseband. A sampling rate of \SI{30}{kSa/s} then covers the full comb, and $2^{20}$ samples correspond to a record length of \SI{35}{\second}. The resulting resolution of $\SI{30}{kSa/s}/2^{20} \approx \SI{30}{\milli\hertz}$ lies well below the linewidths of all comb lines, ensuring that the comb measurement is resolved. [Fig.~\ref{fig:remsat}].

\textit{Phase-space rotation}---The detection chain imparts a constant phase offset on the signal, mainly through the propagation delay in the cables. We compensate for it by rotating the measured quadratures by \SI{-0.386 \pm 0.004}{\radian} in all phase-space plots [Figs.~\ref{fig:semduf}(b) and~\ref{fig:per}], determined by minimizing the distance between the measured points and the theoretical limit cycle. Such constant rotation leaves the shape of the trajectories unchanged and only aligns the measured frame with the theoretical one, in which the drive acts along $Q$ [Eq.~\eqref{eq:heff_main}].

\textit{Global lineshape fit}---Least-squares fitting is not suited for power spectra, since the fluctuations of each spectral bin scale with its mean. For Gaussian noise in the time domain, the real and imaginary parts of each Fourier component are Gaussian. Since we perform no spectral averaging, the power in each bin follows a $\chi^2_2$ distribution, i.e., an exponential distribution~\cite{Brockwell1991}. While a Hann window was applied to the data, the resulting correlation between adjacent spectral bins has a negligible effect given our high frequency resolution relative to the broad lineshapes fitted here. We therefore maximize the Whittle likelihood~\cite{Whittle1953,Myung2003,Roume2023}, which models the measured power $y_k$ in bin $k$ as exponentially distributed with mean $S_k$ given by the model spectrum plus a constant noise floor. This amounts to maximizing
\begin{equation}
    \ln L = - \sum_k \left[ \ln S_k + \frac{y_k}{S_k} \right]
\end{equation}
with respect to $\mathcal{D}$ and $\delta$, jointly for all comb lines. We include only bins more than \SI{5}{\dB} above the estimated noise floor, so that the noise floor does not bias the estimates of $\mathcal{D}$ and $\delta$.

\textit{Linewidth extraction}---We extract the linewidth of each comb line in four steps. 
First, we determine the tails of the comb lines, which carry a substantial part of the peak area but disappear in the noise floor. Far from the line center, the OU lineshape decays as the inverse fourth power of the distance to the center  (see Eq.~(\ref{eq:fapprox}) of the Supplemental Material~\cite{Note1}). 
Figure~\ref{fig:reconshape} plots the positive half of the measured data of two different comb lines ($m=-1$ and $-5$) on a log-log scale in gray. The red solid and blue dashed lines represent the OU theoretical lineshape and the asymptotic  $(\omega-\omega_\mathrm{d} + m \nu_c)^{-4}$ tail, respectively. 
Where the OU lineshape and its asymptotic approximation agree to within \SI{0.5}{dB}, we replace the data by the asymptotic tail. The resulting hybrid lineshape is included in Fig.~\ref{fig:reconshape} in black.
Second, we estimate the noise floor of each line as the median difference between the data (gray) and the asymptotic tails (blue) in the noise-dominated region of the spectrum where the noise floor of the experiment exceeds the asymptotic tail by more than an order of magnitude. The obtained noise floor is included in Fig.~\ref{fig:reconshape} as a green dotted line. 
Third, we determine the peak height of each comb line, which is set by the oscillator's strength $|z_m|^2$ in Eq.~\eqref{eq:psd_ksi}. The global fit uses normalized lines, so that $\mathcal{D}$ and $\delta$ are its only free parameters. We therefore obtain the strengths from a second fit of Eq.~\eqref{eq:psd_ksi}, with $\mathcal{D}$ and $\delta$ fixed to the values of the global fit. The noise floors and strengths are listed in Table~\ref{tab:fitted_zm2_and_nf} of the Supplemental Material~\cite{Note1}.
Finally, we obtain the peak area as the sum of the noise-floor-corrected central part, integrated with the trapezoidal rule, and the analytically integrated tails. The integral linewidth is the ratio of the peak area to the peak height. The experimentally determined linewidths depicted in Fig.~\ref{fig:linewidth} result from averaging over $100$ measurements, and shown along with their standard deviation. The values of the linewidths of all comb lines predicted by the theory and extracted from the experiment are compared in Table~\ref{tab:linewidthtab} of the Supplemental Material~\cite{Note1}.

\begin{figure}[]
    \centering
    \includegraphics[]{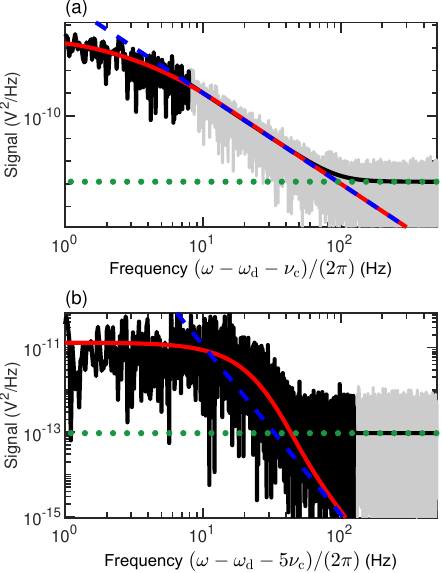}
    \caption{Log-log representation of comb lines for (a) $m=-1$ and (b) $m=-5$. Positive half of the measured peak is plotted as a function of the frequency difference from the center frequency of the comb line (gray). The theoretical OU lineshape obtained from the global fit and its asymptotic tail are included as red solid and blue dashed line, respectively. The reconstructed hybrid OU lineshape is depicted in black. The resulting estimate of the noise floor is shown as a green dotted line.
    }
    \label{fig:reconshape}
\end{figure}

\clearpage
\setcounter{figure}{0}\renewcommand{\thefigure}{S\arabic{figure}}
\setcounter{table}{0}\renewcommand{\thetable}{S\arabic{table}}
\setcounter{equation}{0}\renewcommand{\theequation}{S\arabic{equation}}

\begin{center}
  \textbf{\Large SUPPLEMENTAL MATERIAL}\\[.2cm]
\end{center}

\section{THEORY}

\subsection{Fluctuations of action-angle variables}
We consider a mode with natural vibration frequency $\omega_0$, driven at a frequency $\omega_\mathrm{d}$ with amplitude $F$. In a frame rotating at the drive frequency, the mode is described by the quadratures $Q$ and $P$,
\begin{align}
    \dot Q &= \partial_P g(Q,P) + R_Q(Q,P) + f_Q(t), \label{eq:eom_Q} \\
    \dot P &= -\partial_Q g(Q,P) + R_P(Q,P) + f_P(t), \label{eq:eom_P}
\end{align}
where the rotating-wave-approximated Hamiltonian is
\begin{align}
    g(Q,P) ={}& -\frac{1}{2}\delta\omega(Q^2+P^2) + \frac{M\omega_\mathrm{d}}{8}\alpha_1(Q^2+P^2)^2 \nonumber\\
    &+\frac{(M\omega_\mathrm{d})^2}{24}\alpha_2(Q^2+P^2)^3 \nonumber \\
    &+\frac{(M\omega_\mathrm{d})^3}{64}\alpha_3(Q^2+P^2)^4-\frac{F}{2M\omega_\mathrm{d}}Q.
    \label{eq_app_function_g}
\end{align}

Here, $\delta\omega =\omega_\mathrm{d} - \omega_0$ denotes the detuning and $\alpha_i$ are the nonlinear coefficients of the rotating-wave Hamiltonian. For a nonlinear potential, the coefficients $\alpha_i$ can be related to the corresponding microscopic nonlinearities. For example, for a Duffing nonlinearity with coefficient $\gamma$, one finds $\alpha_1\propto\gamma$. Here we include all three $\alpha_1$, $\alpha_2$ and $\alpha_3$, which are necessary to describe the backbone curve of our resonator \cite{Kallergi2026}. In general,  the coefficients $\alpha_i$ are renormalized by other nonlinearities, including nonlinear coupling to other modes \cite{Dykman1971,Ochs2022,Kallergi2026}. The terms $R_Q$ and $R_P$ describe dissipation. In the absence of the drive, they reduce to $R_Q=-\Gamma_0 Q$ and $R_P=-\Gamma_0 P$. The associated fluctuating forces satisfy the fluctuation-dissipation relation \cite{Dykman1971}
\begin{align}
    \langle f_i(t)f_j(t')\rangle =2D\delta(t-t')\delta_{ij},
\end{align}
where $D=\Gamma_0 k_BT/(M\omega_\mathrm{d}^2)$ is the diffusion coefficient and $i, j \in \{Q, P\}$.

Under resonant driving, nonlinear friction generates higher-order dissipative terms that can give rise to a limit cycle in the rotating frame. A simple phenomenological form of such a friction force is RIFF \cite{Dykman2019},
\begin{align}
    F_\mathrm{RIFF} &= -\eta_\mathrm{RIFF} [F \cos(\omega_\mathrm{d} t)  \dot q(t)]_\mathrm{av} q(t),
\end{align}
where the averaging is over one period $2\pi/\omega_\mathrm{d}$. In the rotating frame, accounting for this nonlinear friction, the dissipative terms become
\begin{align}
    R_Q &= - \Gamma_0 Q + \frac{\eta_\mathrm{RIFF}}{4M} F P^2, & R_P &= -\Gamma_0 P - \frac{\eta_\mathrm{RIFF}}{4M} FQP.
\end{align}

In the absence of noise, these nonlinear dissipative terms can prevent $Q$ and $P$ from relaxing to a fixed point, instead giving rise to closed trajectories corresponding to a limit cycle. 

For weak noise, the motion remains approximately periodic. A remarkable feature of these vibrations is that once the driving exceeds the critical value for the onset of the vibrations, they quickly become strongly non-sinusoidal, which underlies the onset of the frequency comb. 
The vibrations are conveniently described in terms of action-angle variables, $I = (2\pi)^{-1}\oint P\,dQ$ and $\psi = (\partial/\partial I)\int^Q P(I,Q')\,dQ'$ for a system with Hamiltonian $g$ \cite{Landau1982}. In the absence of dissipation and noise, the action $I$ is conserved, while the phase $\psi$ increases linearly in time with the vibration frequency $\Omega(I)=\partial g/\partial I$. The original variables can then be expressed as $Q(I,\psi)$ and $P(I,\psi)$, which are $2\pi$-periodic functions of $\psi$. In the presence of dissipation and fluctuations, the action is no longer conserved, and the dynamics of the phase variable are modified such that
\begin{align}
\label{eq_app_I_phi}
    \dot{I} &= (\partial_\psi Q) R_P - (\partial_\psi P) R_Q + (\partial_\psi Q) f_P - (\partial_\psi P) f_Q,\\
    \dot{\psi} &= \Omega(I) + (\partial_I P) R_Q - (\partial_I Q) R_P + (\partial_I P) f_Q - (\partial_I Q) f_P \,.\nonumber
\end{align}

In the present system, the vibration frequency $\Omega$ is large compared with both the decay rate $\Gamma_0$ and the diffusion rate \cite{Ochs2022}. Therefore, within a single vibration period, the action $I$ changes only slightly, while $\psi$ accumulates approximately $2\pi$. The long-term dynamics can thus be described by averaging Eq.~\eqref{eq_app_I_phi} over one vibration period, yielding \cite{Miller2021}
\begin{align}
\dot I &\approx \overline{R_I(I)} + f_I(I,t),\\
\dot \psi &\approx \Omega(I) + \overline{(\partial_I P) R_Q - (\partial_I Q) R_P} + f_\psi(I,t).
\label{eq:averaged_dynamics}
\end{align}
Here, the overline indicates averaging of the periodic functions over $\psi$ for a given $I$. The noise terms are given by $f_I(I,t) = \{(\partial_\psi Q) f_P - (\partial_\psi P) f_Q\}_\psi$ and $f_\psi(I,t) = \{(\partial_I P) f_Q - (\partial_I Q) f_P \}_\psi$ where the curly brackets denote an average of the periodic coefficients over $\psi$, such that the correlators of the noise terms depend only on the action $I$. In the absence of fluctuations, the stationary orbit $I=I_\mathrm{c}$ is determined by the condition $\overline{R_I} = 0$, with
\begin{align}
\overline{R_I} &= \overline{(\partial_\psi Q) R_P - (\partial_\psi P) R_Q}\\
&= \frac{1}{2\pi}\iint_{\mathcal{A}(g) } dQ\,dP\left(\partial_Q R_Q+\partial_P R_P\right),
\end{align}
where the integral is taken over the interior of the orbit $g(Q,P)=g$. The corresponding vibration frequency of the limit cycle is $\nu(I_\mathrm{c})$, where $\nu(I) = \Omega(I) + \overline{(\partial_I P) R_Q - (\partial_I Q) R_P}$.

In the following, we assume that the noise is sufficiently weak such that fluctuations around the stationary value of the action remain small, allowing us to expand the dynamics around $I=I_\mathrm{c}$. The equations of motion can then be written as
\begin{align}
\delta \dot{I} &= -\alpha(I_\mathrm{c}) \delta I + f_I(I_\mathrm{c},t),\\
\dot{\psi} &= \nu(I_\mathrm{c}) + \beta(I_\mathrm{c}) \delta I + f_\psi(I_\mathrm{c},t),
\end{align}
where $\alpha(I) = -\partial_I \overline{R_I}$ and $\beta(I) = \partial_I \nu(I)$. For a stable limit cycle, we require $\alpha(I_\mathrm{c})>0$. The action variable therefore experiences random kicks from the fluctuating force $f_I$ while relaxing back to its stationary value $I_\mathrm{c}$ at the rate $\alpha(I_\mathrm{c})$. In contrast, $\psi$ increases at the rate $\nu(I_\mathrm{c})$ while being influenced by fluctuations. These fluctuations have two distinct contributions. On the one hand, phase fluctuations act directly on the phase variable through $f_\psi$. On the other hand, fluctuations in the action $I$ are converted into fluctuations of $\psi$ through the nonlinear dependence of the vibration frequency on the action. Specifically, the action-dependent frequency allows fluctuations in the action variable to propagate to the phase variable through the term $\beta(I_\mathrm{c})$. For a weakly nonlinear system, the contribution is typically small. In the strongly nonlinear system considered in this work, however, it can become significant. At the same time,  the direct contribution $f_\psi$ is small for large-amplitude self-sustained vibrations.   Therefore, in our system it is the contribution of the noise that comes from the amplitude fluctuations that becomes the dominating mechanism of the frequency noise.

In the following, we are interested in the effect of frequency fluctuations and the resulting phase diffusion on the spectral shape. For that we consider dynamics of the form
\begin{align}
\dot \psi &= \nu_\mathrm{c} + \zeta(t), & \nu_\mathrm{c} &\equiv \nu(I_\mathrm{c}),
\end{align}
where 
\[\zeta(t)= \beta(I_\mathrm{c})\int^t_{-\infty} dt' e^{-\alpha(I_\mathrm{c})(t-t')} f_I(I_\mathrm{c},t') +f_\psi (I_\mathrm{c},t)
\]
is assumed to be Gaussian distributed.

\subsection{Gaussian phase noise and spectral broadening}
We consider a general stationary Gaussian noise process $\zeta(t)$. Its probability density functional can be written as
\begin{align}
    \mathcal{P}[\zeta(t)] &= 
    \exp\left[ -\frac{1}{2} \int_{-\infty}^{\infty} \int_{-\infty}^{\infty}d t_1 d t_2 \Xi^{-1}(t_1-t_2)\zeta(t_1)\zeta(t_2) \right],
\end{align}
where the function $\Xi$ and its inverse are related by
\begin{align}
    \int_{-\infty}^{\infty} d t_2 \Xi^{-1}(t_1-t_2)\Xi(t_2-t_3) &=\delta(t_1-t_3),
\end{align}
cf. \cite{Feynman1965}. Stationarity requires $\Xi$ to depend only on the time difference. Since the probability functional is Gaussian, the noise correlator is
\begin{align}
    \langle\zeta(t_1)\zeta(t_2)\rangle &= \Xi(t_1-t_2).
\end{align}

One way to characterize different noise processes is in terms of the resulting diffusion, cf. \cite{Anderson1953,*Anderson1954, Kubo1954,*Kubo1957}. For this purpose we define the phase difference $\phi = \psi - \nu_\mathrm{c} t$, which undergoes a random walk $\dot{\phi}=\zeta(t)$. Its mean-square displacement is given by
\begin{align}
    \langle[\phi(t)- \phi(0)]^2\rangle &= \int_0^t\int_0^t d t_1 d t_2 \Xi(t_1-t_2) \nonumber\\ 
    &= 2\int_0^t d\tau (t-\tau)\Xi(\tau),
    \label{eq:MSD_phase}
\end{align}
which is therefore directly related to $\Xi$.

We are interested in the power spectral density of the resonator. In order to calculate it, we expand the complex quadrature $u(t) = Q(t) + i P(t)$ in harmonics of the phase variable,
\begin{align}
    u(t) &= \sum_m z_m(I_\mathrm{c})e^{im\psi(t)}.
\end{align}
We remind that the coordinate $q$ and momentum $p$ of the oscillator are simply related to $u(t)$, namely, $q+i[p/(M\omega_\mathrm{d})] = u(t)\exp(-i\omega_\mathrm{d} t)$. The squared absolute values of the expansion coefficients $|z_m|^2$ determine the oscillator's strength of the different comb lines, which are shifted from the drive frequency $\omega_\mathrm{d}$ by multiples of the limit cycle frequency $m\nu_\mathrm{c}$. For the simple case $\alpha_2 = \alpha_3 = 0$, the trajectories for $I=I_\mathrm{c}$ are given by Jacobi elliptic functions. The double periodicity of these functions allows one to calculate the coefficients $z_m$ analytically \cite{Dykman2019,Ochs2022} (in those papers, $z_m$ was expressed as a function of $g_\mathrm{c}$, where $g_\mathrm{c}$ corresponds to the value of $g$ at $I=I_\mathrm{c}$). We can now write the power spectral density (PSD) of the resonator as
\begin{align}
    S(\omega) &= \frac{1}{4} \int_{-\infty}^\infty d \tau e^{i(\omega-\omega_\mathrm{d})\tau} \langle u(t+\tau) u^\ast(t) \rangle\\
     &= \frac{1}{4} \sum_m |z_m|^2 \int_{-\infty}^\infty d\tau e^{i(\omega- \omega_\mathrm{d} + m \nu_\mathrm{c}) \tau} \langle e^{im [\phi(\tau)-\phi(0)]}\rangle, \nonumber
\end{align}
where we have used that $\psi(t)=\nu_\mathrm{c} t+\phi(t)$. 

For the stationary Gaussian noise process assumed here, the correlator can be related \cite{Feynman1965,Anderson1953,Kubo1954} to the mean square displacement of $\phi$ from Eq.~\eqref{eq:MSD_phase}
\begin{align}
    \langle e^{im [\phi(\tau)-\phi(0)]}\rangle = e^{-m^2 \langle [\phi(\tau)-\phi(0)]^2\rangle /2}.
\end{align}
Inserting this expression into the PSD yields
\begin{align}
    S(\omega) = &\frac{1}{2}\sum_m|z_m|^2 S^{(m)}(\omega-\omega_\mathrm{d}+m\nu_\mathrm{c})\label{eq:PSD_gaussian} \\
    S^{(m)}(\Delta) = &\int_0^\infty d\tau \cos(\tau\Delta) e^{-m^2\int_0^\tau d x(\tau-x)\Xi(x)}.\nonumber
\end{align}
The function $S^{(m)}(\Delta)$ determines the lineshape of the $m$th harmonic of the limit cycle. It is normalized such that $\int d\Delta S^{(m)}(\Delta) =\pi$. The width and shape of $S^{(m)}(\Delta)$ depend explicitly on the harmonic index $m$ and on the specific form of $\Xi(t)$.

\subsubsection{White noise}
For white Gaussian noise, the correlator is

\begin{align}
    \Xi(\tau) &= 2\mathcal{D}\delta(\tau)\,.
    \label{eq:Xi_white}
\end{align}
The parameter $\mathcal{D}$ is the phase diffusion coefficient. From Eq.~\eqref{eq:MSD_phase}, the mean-square phase displacement grows linearly with time \cite{Berstein1938},
\begin{align}
    \langle[\phi(t)-\phi(0)]^2\rangle &=2\mathcal{D}t,
\end{align}
With Eq.~\eqref{eq:PSD_gaussian} taken into account, this yields the PSD of the resonator  
\begin{align}
    S_\mathrm{white}(\omega) 
    &= \frac{1}{2}\sum_m|z_m|^2 \frac{m^2\mathcal{D}} {(\omega-\omega_\mathrm{d}+m\nu_\mathrm{c})^2+(m^2\mathcal{D})^2}.
    \label{eq:WGN}
\end{align}
Each harmonic produces a Lorentzian line $S_\mathrm{white}^{(m)}$ with a half-width $m^2\mathcal{D}$.

\subsubsection{\texorpdfstring{$1/f^\alpha$}{1/f-alpha} noise}
We now compute the power spectral density (PSD) of the resonator for $1/f^\alpha$ noise with $\alpha<1$ in an appreciably broad frequency range. For this noise, the correlation function $\Xi$ decays as a power law in time,
\begin{align}
    \Xi(t) &= 2 \mathcal{D} |t|^{\alpha-1},
    \label{eq:ksi1f}
\end{align}
with the noise strength $\mathcal{D}$. Then, we find that the mean-square phase displacement is given by
\begin{align}
    \langle[\phi(t)-\phi(0)]^2\rangle &=  \frac{4 \mathcal{D}}{\alpha(\alpha +1)} t^{\alpha+1}.
    \label{eq:Diffusion_1_over_f}
\end{align}
Inserting Eq.~\eqref{eq:ksi1f} into the general expression for the PSD, Eq.~\eqref{eq:PSD_gaussian}, gives
\begin{align}
    S_{1/f^\alpha}^{(m)}(\Delta) &=\int_0^\infty d\tau \cos(\tau \Delta) e^{-\frac{2 m^2 \mathcal{D}}{\alpha(\alpha+1)} \tau^{\alpha+1}}.
    \label{eq:PSD_1f}  
\end{align}
%


\subsubsection{Ornstein--Uhlenbeck noise}
We now consider an Ornstein--Uhlenbeck process where the PSD of the noise follows a Lorentzian distribution and
\begin{align}
    \Xi(t) &= \frac{\mathcal{D}}{\delta}e^{-|t|/\delta}.
    \label{eq:Xi_OU}
\end{align}
The mean-square phase displacement of the phase difference $\phi$ evaluates to
\begin{align}
    \langle[\phi(t)-\phi(0)]^2\rangle &=  2\mathcal{D} \left( t + \delta e^{-t/\delta}-\delta\right).
    \label{eq:Diffusion_OU}
\end{align}
To connect explicitly to the main text, we indicate that, for the Ornstein--Uhlenbeck noise, in Eq.~\eqref{eq:psd_ksi} of the main text we have 
\begin{equation}
\int_0^t \Xi(x)(t - x) dx = \mathcal{D} \left( t + \delta e^{-t/\delta}-\delta\right).
\end{equation}
For $t\ll\delta$, we have $\langle[\phi(t)-\phi(0)]^2\rangle\simeq\mathcal{D} t^2 /\delta$, an anomalous diffusion corresponding to ballistic motion. For $t\gg\delta$, the mean-square phase displacement approaches $\langle[\phi(t)-\phi(0)]^2\rangle\simeq2\mathcal{D} t$, corresponding to normal diffusion.

The oscillator PSD is found by inserting Eq.~\eqref{eq:Xi_OU} into Eq.~\eqref{eq:PSD_gaussian}.The resulting integral can be reduced to the incomplete Gamma function
\begin{align}
    S_\mathrm{OU} 
    =&\frac{\delta}{2}\sum_m|z_m|^2 e^{a_m}\mathrm{Re} \{ a_m^{-a_m+ib_m} \gamma(a_m-ib_m, a_m)\}, \label{eq:PSD_OU}
\end{align}
where $a_m = m^2 \delta \mathcal{D}$, $b_m = (\omega-\omega_\mathrm{d}+m\nu_\mathrm{c})\delta $ and $\gamma$ is the lower incomplete Gamma function. 

The parameter $a_m$ compares the correlation time $\delta$ of the noise with the decay time $1/(m^2\mathcal{D})$ of the phase correlator for white noise of the same $\mathcal{D}$. For $a_m\gg 1$, the correlator decays while the phase displacement is still ballistic. Near its maximum, each peak then becomes a Gaussian,
\begin{align}
    S_\mathrm{OU}^{(m)}(\Delta) \simeq \sqrt{\frac{\pi}{2\sigma_m^2}}\, e^{-\Delta^2/2\sigma_m^2}, \qquad \sigma_m^2 = \frac{m^2\mathcal{D}}{\delta},
\end{align}
with a half-width $\sqrt{2\ln 2}\,\sigma_m$ that grows linearly with $m$. In contrast, for $a_m\ll 1$, the phase correlator decays slowly compared to the correlation time. For $|b_m|\ll 1$, the peaks are Lorentzian with half-width $m^2\mathcal{D}$, as for white noise. This is consistent with the noise correlator in Eq.~\eqref{eq:Xi_OU} approaching the white-noise correlator, Eq.~\eqref{eq:Xi_white}, for $\delta\to 0$.
\begin{figure}
    \centering
    \includegraphics[]{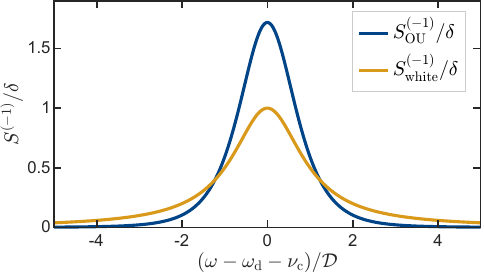}
    \caption{Lineshape of the $m=-1$ comb line near $\omega\approx\omega_\mathrm{d}+\nu_\mathrm{c}$ for Ornstein–Uhlenbeck noise (blue) with $\delta=1/\mathcal{D}$ (i.e. $a_1=1$), compared with white noise assuming the same parameters (dark yellow).}
    \label{fig:app_comparison_spectra}
\end{figure}

In Fig.~\ref{fig:app_comparison_spectra}, we compare the lineshape of the $m=-1$ peak for Ornstein--Uhlenbeck noise (blue) and white noise (dark yellow), with the same $\mathcal{D}$ and $\delta = 1/\mathcal{D}$, i.e., in the intermediate regime $a_1 = 1$. OU noise displays a narrower peak that falls off quickly for larger detunings from the center frequency.
In this regime, the peaks from the Ornstein--Uhlenbeck type noise are significantly non-Lorentzian, and the half-width of the $m$th peak is not proportional to $m^2$. Sufficiently far away from the peak, where the absolute value of the detuning $\omega -\omega_\mathrm{d} + m \nu_\mathrm{c}$ is large compared to $m^2 \mathcal{D}$ and $1/\delta$, we can use partial integration to approximate
\begin{align}
    S_\mathrm{OU}^{(m)}(\Delta) \approx \frac{m^2 \mathcal{D}}{\delta^2 \Delta^4}.
    \label{eq:fapprox}
\end{align}
Therefore, the tails of the peaks fall off as the reciprocal fourth power of the distance to the peak, in contrast to the tails of the Lorentzian peaks that fall off as the reciprocal square of this distance.

\subsection{Parameter values of the effective Hamiltonian describing the experiment}

The trajectories obtained in Fig.~\ref{fig:semduf}(b) are the solution of the flow equations for the coordinates $Q$ and $P$ as given in Eq.~\eqref{eq:eom_Q} and Eq.~\eqref{eq:eom_P} neglecting the noise contributions $f_Q(t)$ and $f_P(t)$. The parameter values used to numerically integrate these trajectories are obtained directly from experimental data. The fundamental bare resonance frequency is $f_0 = \SI{6.52}{\mega\hertz}$ (where $\omega_0 = 2\pi f_0$), with a linear damping rate of $2\Gamma_0/(2\pi) \approx \SI{21}{\hertz}$. The system is driven at a power of \SI{0}{\dBm} with a drive detuning of $\delta\omega = 2\pi \cdot \SI{840}{\hertz}$. 

As the phase space data are measured directly as voltage signals from the lock-in amplifier, the nonlinear coefficients in the flow equations correspond to the physical parameters scaled by mass and transduction, and we use the values $M a^2 \alpha_1 = \SI{107.91}{\per\volt\squared}$, $M^2 a^4 \alpha_2 = \SI{0.576}{\second\per\volt\tothe{4}}$, and $M^3 a^6 \alpha_3 = \SI{0.01}{\second\squared\per\volt\tothe{6}}$. The RIFF damping is estimated as $a^2 \eta_{\text{RIFF}} \approx \SI{-2.70e-3}{\second\per\volt\squared}$, where $a = \SI{3.5181e-5}{\meter\per\volt}$ is the conversion factor between measured and physical units. 

\section{LINEWIDTH FIT DETAILS}
\subsection{Noise floor and peak height of comb lines}
Table~\ref{tab:fitted_zm2_and_nf} summarizes the values of the noise floor and peak strength of each comb line required for the linewidth extraction described in the End Matter. Table~\ref{tab:linewidthtab} compares the theoretical linewidths of the OU comb lines with the experimentally obtained values along with their standard deviation. All experimental data are presented in units of voltage as obtained directly from the measurement instruments to avoid ambiguity regarding the scaling to physical units of displacement.

\begin{table}[hbt!]
\centering
\caption{Noise floor and oscillator's strength $|z_m|^2$ of each comb line $m$.}
\label{tab:fitted_zm2_and_nf}
\begin{minipage}{\columnwidth} 
\begin{ruledtabular}
\begin{tabular}{ccc}
 $m$ & Noise floor (V$^2$/Hz) & $|z_m|^2$ (V$^2$) \\
\hline
3 & \num{8.108e-14}  & \num{2.149e-10} \\
2 & \num{8.612e-14}  & \num{1.085e-08} \\
1 & \num{8.949e-14}  & \num{1.967e-07} \\
- 1 & \num{1.0412e-13} & \num{5.457e-07} \\
- 2 & \num{1.074e-13}  & \num{1.361e-07} \\
- 3 & \num{1.000e-13}  & \num{1.970e-08} \\
- 4 & \num{9.001e-14}  & \num{2.330e-09} \\
- 5 & \num{8.776e-14}  & \num{2.414e-10} \\
 \end{tabular}
\end{ruledtabular}
\end{minipage}
\end{table}

\begin{table}[htb]
\caption{Theoretical versus experimental linewidths (mean of $100$ measurements and standard deviation).}
\label{tab:linewidthtab}
\begin{ruledtabular}
\begin{tabular}{cccc}
$m$ & Theory (\unit{\hertz}) & Measured (\unit{\hertz}) & Std.\ dev.\ (\unit{\hertz}) \\
\hline
3 & \num{16.006} & \num{16.777} & \num{0.959} \\
2 & \num{9.468}  & \num{9.794}  & \num{0.542} \\
1 & \num{3.292}  & \num{3.528}  & \num{0.284} \\
- 1 & \num{3.292}  & \num{3.529}  & \num{0.284} \\
- 2 & \num{9.468}  & \num{9.777}  & \num{0.541} \\
- 3 & \num{16.006} & \num{16.405} & \num{0.926} \\
- 4 & \num{22.577} & \num{23.134} & \num{1.177} \\
- 5 & \num{29.149} & \num{30.658} & \num{2.009} \\
\end{tabular}
\end{ruledtabular}

\end{table}

\section{ADDITIONAL EXPERIMENTAL DATA}

\subsection{Signature of the frequency comb in the response curve}
\begin{figure}
    \centering
    \includegraphics[]{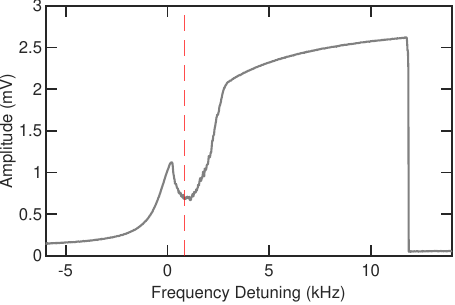}
    \caption{Frequency response of the single nanomechanical mode at a drive power of \SI{0}{\dBm}. The characteristic dip in the otherwise nonlinear response heralds the formation of the frequency combs related to the vibrations in the rotating frame that cause a reduction in the amplitude of the fixed vibrations. The red dashed line marks the offset of the drive frequency from the bare resonance by \SI{840}{\hertz} used in the comb measurements of the main text.}
    \label{fig:duf}
\end{figure}

Figure~\ref{fig:duf} shows the frequency response at the drive power of \SI{0}{\dBm} used for the comb measurements in the main text. Its overall shape is Duffing-like, with higher-order nonlinearities contributing at this drive~\cite{Kallergi2026}. Over a range of frequencies, the response amplitude exhibits a pronounced dip, as also reported in Ref.~\cite{Ochs2022}. The dip marks the onset of the frequency comb: the lock-in amplifier records only the component at the drive frequency, and spectral weight is redistributed from this central peak to the comb lines. The dip thus identifies the drive frequencies at which the comb exists. Its deepest point, at $\delta\omega/(2\pi) = \SI{840}{\hertz}$, coincides with the strongest comb, i.e., the one with the most and largest comb lines. We therefore use this offset throughout the main text.

\subsection{Lineshape fits of the remaining comb lines}
\begin{figure}
    \centering
    \includegraphics[]{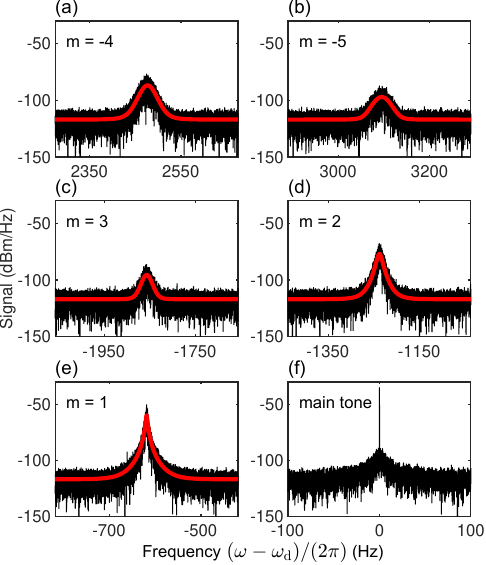}
    \caption{(a)--(e) Close-ups of the comb lines of the comb in Fig.~\ref{fig:semduf}(c) not shown in Figs.~\ref{fig:semduf}(d)--(f) of the main text, including both the measured spectra (black) and global OU fit (red) for $m = -4$, $-5$, $3$, $2$, and $1$. (f) The main tone (forced vibration at the drive frequency) is limited by the spectral resolution of \SI{30}{\milli\hertz}.
    }
    \label{fig:remsat}
\end{figure}

Figure~\ref{fig:semduf}(c) of the main text shows the full comb together with the global OU fit, and Figs.~\ref{fig:semduf}(d)--(f) show close-ups of the lines $m = -1$, $-2$, and $-3$. For completeness, Fig.~\ref{fig:remsat}(a)--(e) shows close-ups of the remaining lines, $m=-4$, $-5$, $3$, $2$, and $1$, together with the same global fit yielding 
 $\mathcal{D} = \SI{8.0 \pm 0.3}{\radian\squared\per\second}$ 
and $\delta = \SI{29.5 \pm 0.9}{\milli\second}$. 

The fit excellently reproduces all measured comb lines. Figure~\ref{fig:remsat}(f) displays the main tone directly at the drive frequency, i.e. the forced vibrations at $\omega_\mathrm{d}$. It is significantly narrower than the comb lines, since its phase is locked to the drive and thus not subject to phase diffusion. Its width is limited by the spectral resolution of \SI{30}{\milli\hertz}.

\subsection{Obtaining the frequency comb from the phase space trajectory}
\begin{figure}
    \centering
    \includegraphics[]{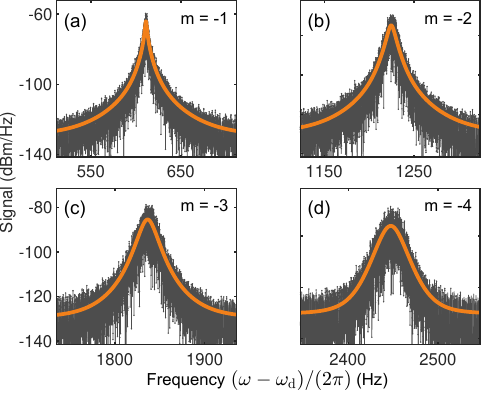}
    \caption{Exemplary comb lines of the frequency comb obtained from the Fourier transformation of the phase space trajectory of the limit cycle in Fig.~\ref{fig:per}(a) of the main text. The global OU fit (orange lines) yields $\mathcal{D} = \SI{5.6 \pm 0.2}{\radian\squared\per\second}$ and $\delta = \SI{25.4 \pm 0.5}{\milli\second}$.}
    \label{fig:whittlepert}
\end{figure}

The comb can also be obtained from the measured phase-space trajectory. We perform a Fourier transform on the stationary part of the trajectory (orange) in Fig.~\ref{fig:per}(a), after the system has settled on the limit cycle. The resulting spectrum shows comb lines spaced by $\nu_\mathrm{c}/(2\pi) = \SI{611.7}{\hertz}$, and Fig.~\ref{fig:whittlepert} displays the lines $m = -1$ to $-4$. Applying the same global lineshape fit as in the main text (orange lines) yields 
$\mathcal{D} = \SI{5.6 \pm 0.2}{\radian\squared\per\second}$ and $\delta = \SI{25.4 \pm 0.5}{\milli\second}$. This correlation time agrees with the relaxation time $\SI{25 \pm 6}{\milli\second}$ obtained from the Poincar\'e analysis of the same data.

The two data sets that we present were recorded many months apart. Over this period, the limit-cycle frequency shifted from \SI{618.9}{\hertz} to \SI{611.7}{\hertz}, i.e., by only \SI{1.2}{\percent}, which demonstrates the long-term stability of the system. The associated shift in the quasienergy accounts for the modified phase diffusion coefficient $\mathcal{D}$.

\subsection{Lineshapes of spectral peaks in the stable regime}

\begin{figure}
    \centering
    \includegraphics[]{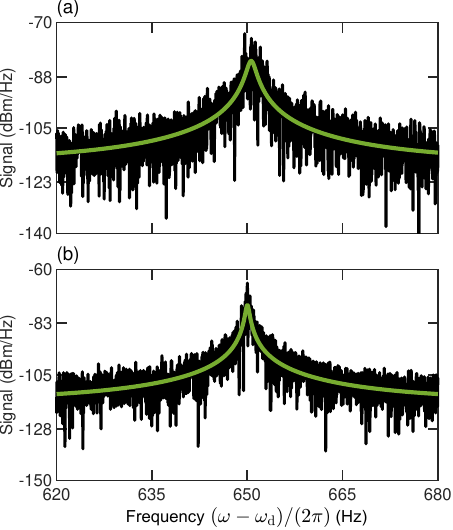}
    \caption{Peak in the spectrum ($m=-1$) of the driven nonlinear resonator in the stable regime (before limit cycle formation). (a) Spectral peak measured at a drive power of \SI{0}{\dBm} with the drive frequency detuning from the bare resonance by \SI{0}{\hertz}. Green line depicts Lorentzian fit. (b) Spectral peak measured at the same drive power with a drive frequency offset from the bare resonance of \SI{70}{\hertz}. Green line depicts Lorentzian fit yielding a narrower linewidth.}
    \label{fig:loron}
\end{figure}

As described in the main text, the primary noise responsible for phase diffusion is OU noise. However, in the regime right before the onset of the frequency comb, thermal white Gaussian noise is dominating. In this regime, the spectrum is described by two spectral peaks equidistantly spaced from the drive frequency~\cite{Huber2020}. In Fig.~\ref{fig:loron}(a), the Lorentzian fit of the $m=-1$ spectral peak measured for a drive frequency detuning from the bare resonance frequency by \SI{0}{\hertz} and a drive power of \SI{0}{\dBm} is depicted. The fit yields a linewidth of \SI{1.36}{\hertz}. With increasing detuning frequency, the system approaches the Hopf bifurcation and the linewidth of the peaks decreases~\cite{Ochs2022}. This is illustrated in  Fig.~\ref{fig:loron}(b) for the peak at a drive frequency detuning from the bare resonance frequency by \SI{70}{\hertz} and a drive power of \SI{0}{\dBm}. The Lorentzian fit yields a reduced linewidth of \SI{0.58}{\hertz}.

\section{NOISE MODEL COMPARISON}
\subsection{White Gaussian noise}
\begin{figure}[]
    \centering
    \includegraphics[]{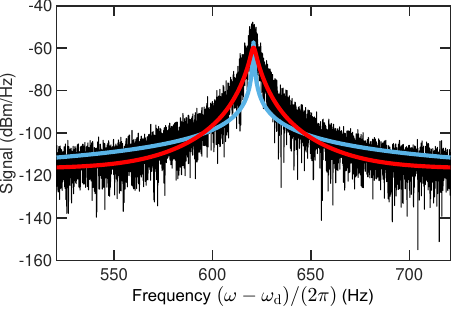}
    \caption{Comparison of OU noise and white Gaussian noise as the origin of the phase diffusion broadening the comb lines. Data of the $m=-1$ comb line is shown along with its global OU fit (red) and a Lorentzian fit according to Eq.~\eqref{eq:WGN} for white Gaussian noise (light blue).}
    \label{fig:wrong_lor}
\end{figure}

White Gaussian noise yields the comb spectrum of Eq.~\eqref{eq:WGN}, with Lorentzian lines whose widths grow as $m^2$. First, the measurement of the linewidths presented in Fig.~\ref{fig:linewidth} of the main text does not yield the expected $m^2$ scaling. 
Second, Fig.~\ref{fig:wrong_lor} shows the $m=-1$ comb line along with the global OU fit (red) and a Lorentzian fit (light blue). Clearly, the Lorentzian fails to describe the measured lineshape.

\subsection{\texorpdfstring{$1/f^\alpha$}{1/f-alpha} Gaussian noise}

\begin{figure*}[]
    \centering
    \includegraphics[]{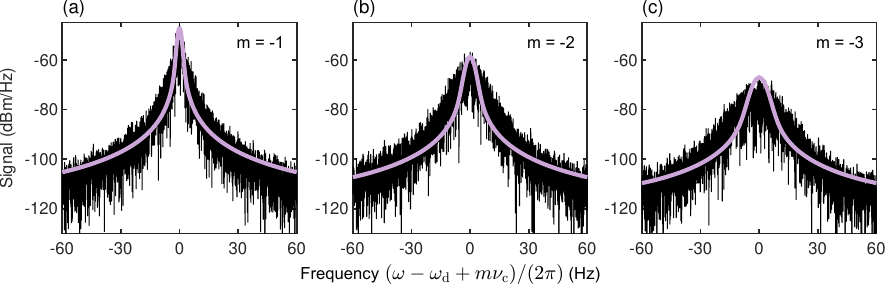}
    \caption{Comparison of comb lines with theory for $1/f^\alpha$ Gaussian noise as the origin of the phase diffusion broadening the comb lines. (a)--(c) Comb lines with $m=-1,-2,-3$ along with the theoretically predicted lineshape in the presence of $1/f^\alpha$ noise with $\mathcal{D} = 0.2~$$\mathrm{rad}^2\,\mathrm{s}^{-(\alpha+1)}$ and $\alpha = 0.8$ (violet).}
    \label{fig:wrong_sats}
\end{figure*}

Gaussian noise with a $1/f^\alpha$ spectrum has also been observed in micro- and nanomechanical systems~\cite{Fong2012,Sun2016}. We therefore test whether it can explain the observed phase diffusion. The noise correlation function of $1/f^\alpha$ noise can be described by Eq.~\eqref{eq:ksi1f}. Figure~\ref{fig:wrong_sats} compares the measured $m=-1$, $-2$ and $-3$ comb lines with the theoretical expectation for $1/f^\alpha$ noise. For this, we solve the integral in Eq.~\eqref{eq:psd_ksi} for the noise correlation function Eq.~\eqref{eq:ksi1f} for a value of $\mathcal{D} = 0.2$ and a dimensionless exponent $\alpha = 0.8$. We note that $\mathcal{D}$ carries units of $\mathrm{rad}^2\,\mathrm{s}^{-(\alpha+1)}$. The theory can be adjusted to agree with the $m=-1$ comb lineshape. However, it does not describe the subsequent comb lines for $m=-2$ or $m=-3$. We therefore conclude that $1/f^\alpha$ Gaussian noise can be excluded as a potential source of the phase diffusion.

\clearpage
\bibliography{Bibliography}

\end{document}